\documentclass[journal,letterpaper]{IEEEtran_nssmic}

\usepackage{graphicx}  

\usepackage{amssymb}
\begin{document}
%
\title{Beam Tests of Ionization Chambers \\ for the NuMI Neutrino Beam}
%
%

\author{Robert~M.~Zwaska, James~Hall, Sacha~E.~Kopp, Huican~Ping, Marek~Proga, 
        Albert~R.~Erwin, Christos~Velissaris, Deborah~A.~Harris, Donna~Naples, 
        Jeffrey~McDonald, David~Northacker, Milind~Diwan, and Brett~Viren
\thanks{Manuscript received December 2, 2002;
       This work was supported by the U.S. Department of Energy, DE-AC02-76CH3000
       and DE-FG03-93ER40757, DE-FG02-05ER40896, and the Fondren Foundation.}  
\thanks{R.~M.~Zwaska, J.~Hall, S.~E.~Kopp, and M.~Proga are with the Department of
        Physics, University of Texas at Austin,
        Austin, TX 78712 USA (e-mail: zwaska@mail.hep.utexas.edu;
        kopp@mail.hep.utexas.edu).}
\thanks{A.~R.~Erwin, H.~Ping and C.~Velissaris are with the Department of 
	Physics, University of Wisconsin at Madison, Madison, WI 53706 USA.}
\thanks{D.~A.~Harris is with the Fermi National Accelerator Laboratory, Batavia,
        IL 60510 USA.}
\thanks{D.~Naples, J.~McDonald, and D.~Northacker are with the Department of Physics,
        University of Pittsburgh, Pittsburgh, PA 15260 USA.}
\thanks{M.~Diwan and B.~Viren are with the Brookhaven National Laboratory, Upton,
        NY 11973 USA.}}
\maketitle

\begin{abstract}
We have conducted tests at the Fermilab Booster of ionization chambers to 
be used as monitors of the NuMI neutrino beamline.  The chambers were 
exposed to proton fluxes of up to 10$^{12}$ particles/cm$^2$/1.56$\mu$s.  We studied  
space charge effects which can reduce signal collection from 
the chambers at large charged particle beam intensities.
\end{abstract}


\section{Introduction}
%
%
%
%

\PARstart{T}{he} Neutrinos at the Main Injector (NuMI) beamline at the Fermi
National Accelerator Laboratory \cite{numi} will generate an intense
$\nu_{\mu}$ beam from the decays of mesons produced in the 
collisions of 120~GeV protons in a graphite target.  The mesons 
are focused by magnetic ``horns'' into a 675~m evacuated volume to 
allow decays to neutrinos.  A downstream Aluminum/Steel absorber and
bedrock absorb the remnant hadrons and muons in the beam, leaving only 
neutrinos.  The facility is expected deliver beam to neutrino
experiments, beginning with the MINOS neutrino oscillation experiment, 
starting in early 2005.

The meson decays
$\pi/K \rightarrow \mu \nu_{\mu}$ produce an energetic muon for every
neutrino, allowing monitoring and validation of the neutrino beam 
focusing to be accomplished by monitoring of the muon flux.  As in 
several previous experiments, 
the muon flux and remnant hadron flux at the end of the decay volume
will be measured by arrays of ionization chambers
\cite{blair,russians,k2k,cern}.

The beam monitoring system will measure the intensity and spatial 
distribution of the hadron beam
at the end of the decay tunnel, upstream of the absorber, and of the muon beam after
the absorber and at several stations in the bedrock. The monitoring system
will consist of 2~m $\times$ 2~m arrays of ionization chambers with 25~cm 
inter-chamber spacing, with one chamber array in each of
the above stations.  The hadron and muon fluxes are measures of 
any targeting or focusing failures.  The peak charged particle fluxes 
in one 9 $\mu$s accelerator burst will be 2000, 25, 3, and $1.5\times10^6$/cm$^2$ 
in the four monitoring stations.

\begin{figure}[t]
\centering
\includegraphics[width=3.5in]{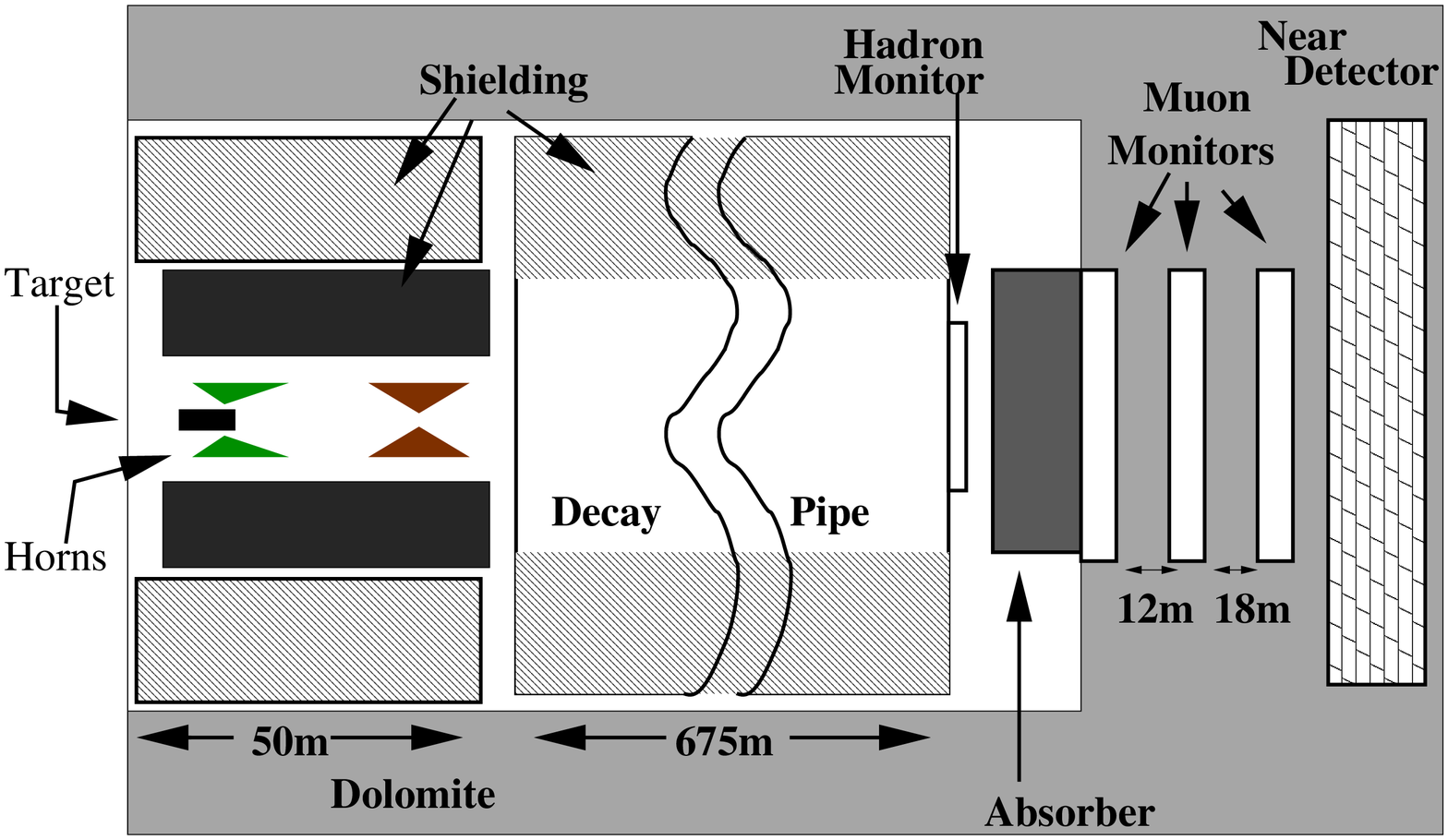}
\caption{Pictorial diagram of the NuMI beamline.  The 120~GeV proton beam is incident
on the target producing a hadron beam.  The positive hadrons are focused by the
horns, of which the pions travel into the decay pipe where they decay into muons
and muon neutrinos, making the neutrino beam.  The hadron monitor and muon
monitors measure hadron and muon fluxes at their locations and are constructed 
of the ion chambers described herein.}
\label{fig_beamline}
\end{figure}

Each ionization chamber will measure the flux of charged particles by using
an applied electric field to collect the ionization created in a helium gas volume.
The charge measured from each chamber will be proportional to the charged particle
flux at that location. By using an array of chambers the spatial distribution of
beam intensity can be inferred.  While operated without gas amplification 
the signal from the intense NuMI beam in one 8~cm $\times$ 8~cm ionization chamber
will be 33000, 1400, 170, and 83~pC at the four stations.

The individual ionization chambers within each array are parallel plate
chambers made up of two
4'' square ceramic plates with Ag-Pt electrodes.   One plate has a single
electrode that applies HV bias.  The second plate has two electrodes: 
a central square sense pad measuring 3''~$\times$~3'',
surrounded by a 1~cm guard ring.  The sense pad is connected into the electronics
which provides a virtual ground.  The guard ring is grounded. The chamber gas is pure
helium at atmospheric pressure.

The major limitation of ionization chambers used as beam monitors has
been 
space charge build up inside the chamber.  Intense particle fluxes
release 
sufficient ionized charge in the chamber gas so as to create a reverse 
electric field inside the chamber.  With the
reduced  net field in the chamber, ions require a longer time to reach the
collection  electrodes, and hence suffer more recombination loss in the gas.  
Recombination losses increase at larger particle fluxes, resulting in a 
non-linear performance of the ionization chamber beam monitor at large 
intensities.


\section{Booster Beam Test}

A beam test of prototype ionization chambers was undertaken at the Fermilab
Booster accelerator, which delivers up to 10$^{12}$~protons/cm$^2$ of 8~GeV in a 
1.56~$\mu$sec spill.  Tested were two chambers, one with a 1 mm electrode 
spacing, the other with 2~mm, with continuos gas flow.  We studied the shape of the 
ionization vs. voltage plateau curve at several intensities 
and the linearity of the chamber response 
vs. beam intensity at several applied voltages.

For our beam test we placed two ionization chambers in the beamline.  The chambers 
were housed in a stainless steel vessel with .005''~Ti beam entrance and exit windows.
Electrical feedthroughs were made with stainless steel compression fittings and PEEK 
plastic.

Two gas mixtures were used in the beam test.  The primary gas was pure helium supplied from 
a cylinder with 99.998\% purity. The other gas was a mixture consisting of 98\% Helium
and 2\% Hydrogen, with $\le$ 20 p.p.m. of impurities.  The gas was first passed through 
a getter and gas analyzer.  Online measurement indicated impurites of $<$~1.5~p.p.m.
All gas seals were metallic consisting of compression fittings or copper gaskets
compressed by vacuum fittings.

Upstream of our ion chamber vessel Fermilab provided  a secondary
emission monitor (SEM), to locate the beam.  The SEM provided targeting information
upstream of the chamber, but was not usable in the analysis because the beam
diverged after passing through the SEM, and it was only capable of measuring either
the horizontal or vertical profile at any given time.  Furthermore, the SEM was
insensitive below $2~\times~10^{11}$ protons per pulse.  During the portion of
datataking above  $2~\times~10^{11}$ the SEM indicated constant spot size.

A beam toroid was the primary method of
measuring the beam intensity delivered to the apparatus.  The toroid was supplied
with an amplifier and ADC whose least significant bit was $5\times10^9$ protons.

\begin{figure}[t]
\centering
\includegraphics[width=3.4in]{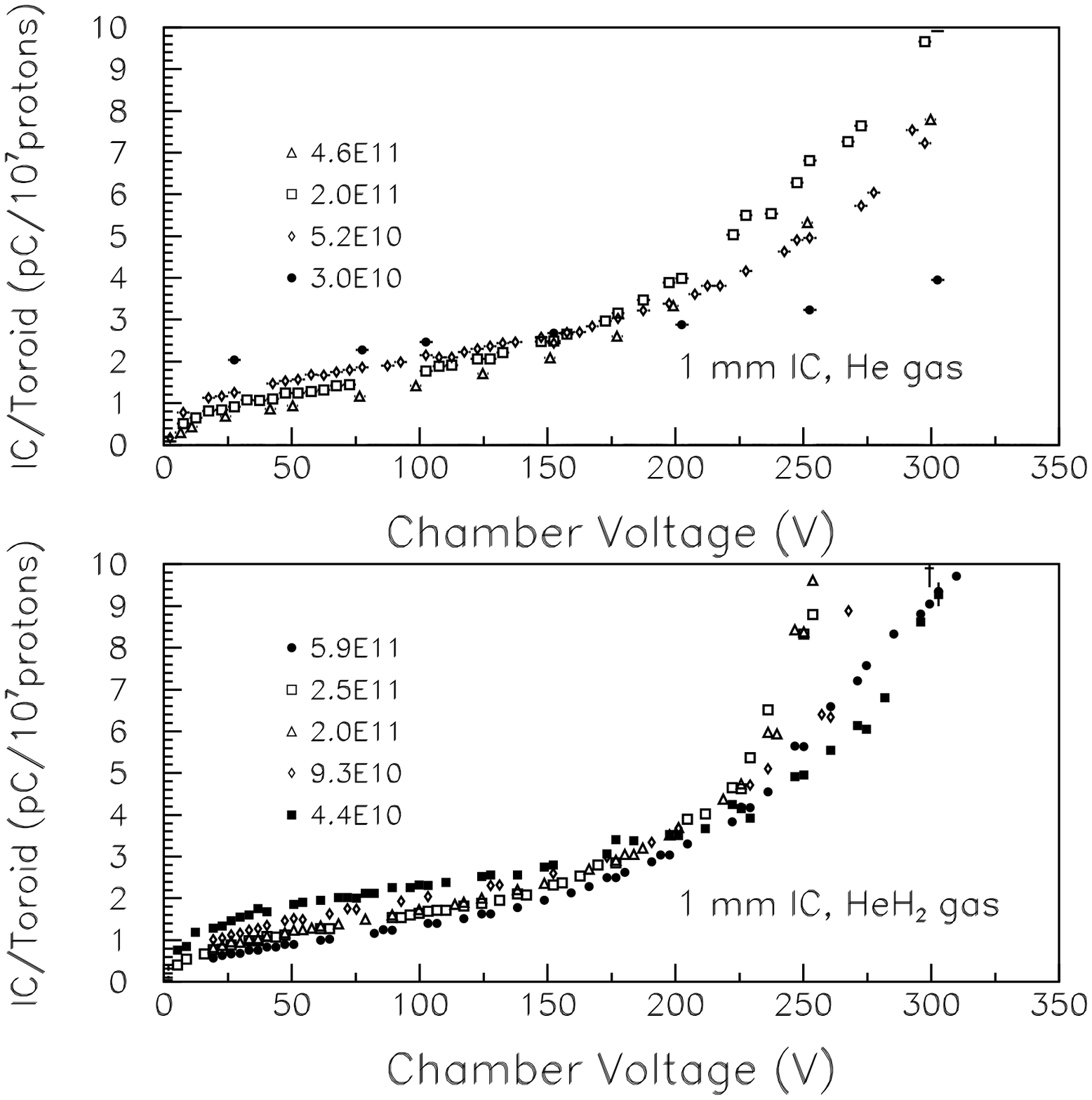}
\caption{High voltage scans of the 1~mm ion chamber (IC) in 
Helium  and Helium-Hydrogen at various beam intensities (noted
in units of protons/spill).    
The vertical axis is the ratio of charge collected from the IC
to the beam intensity measured by the toroid.  Each point is the average
of 10-20 beam spills.}
\label{fig_hv1mm}
\end{figure}

Rigidly attached on the outside of the ion chamber 
vessel were two beam profile chambers
fashioned out of G-10 circuit board and epoxy.  Each chamber had a segmented
signal electrode composed 
of 1$\times$10~cm$^2$ strips.  One chamber provided the vertical profile, the other
the horizontal profile.  The profile chambers were the primary method for determining 
beam size.  Gas flow was the same as that passed through the vessel.  The profile
chambers indicated constant beam spot size $\sim$5~cm$^2$.  

The signal from each of the ion chambers and beam profile chambers were read out into
a charge integrating amplifier and then into an ADC \cite{SWIC}. The electronics 
were triggered on an accelerator clock signal shortly before the beam pulse.  The 
charge integration time could be altered and taken between beam spills, allowing
measurement of pedestals and backgrounds.

\section{Experimental Results}

\begin{figure}[t]
\centering
\includegraphics[width=3.4in]{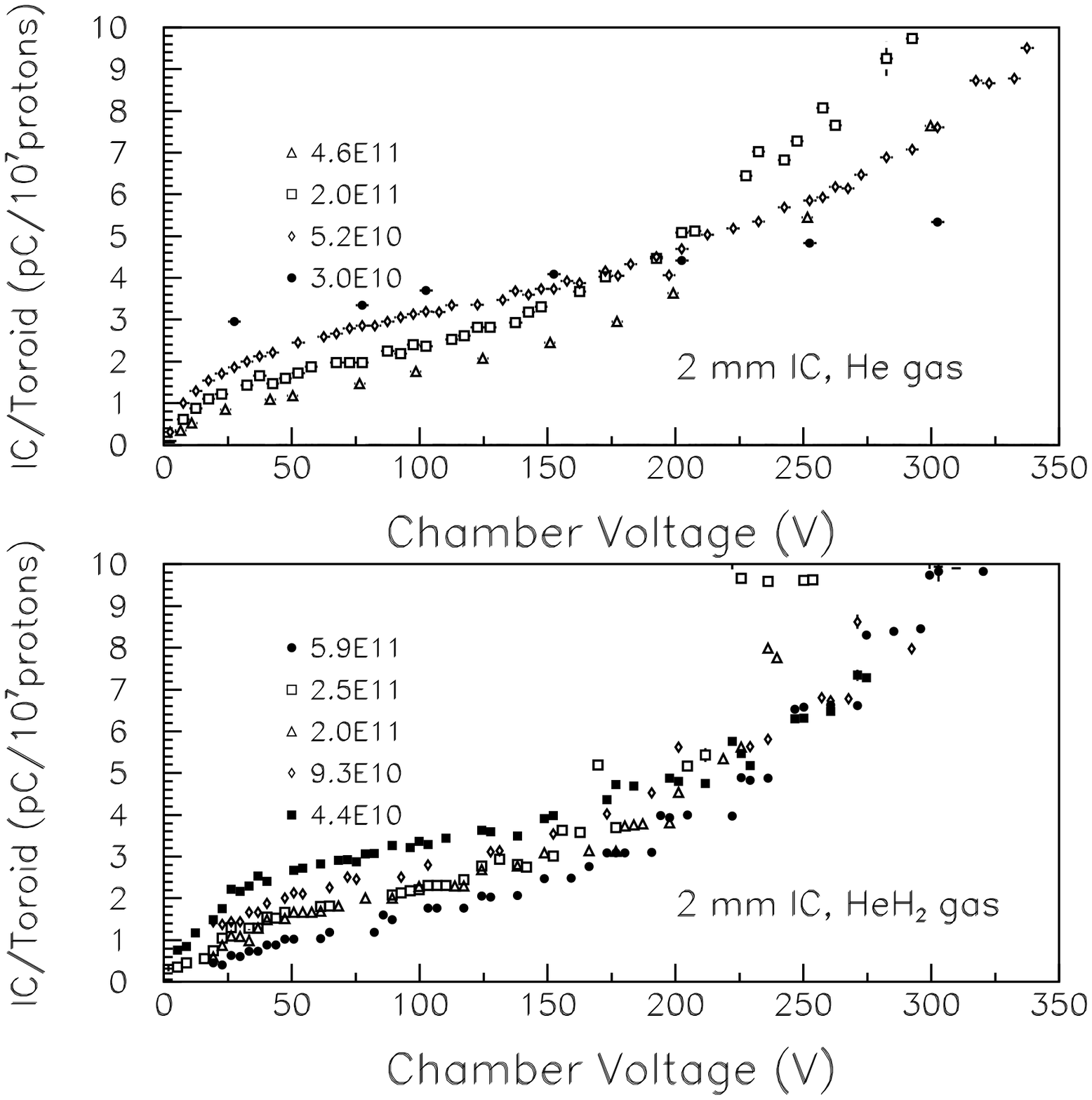}
\caption{High voltage scans of the 1~mm ion chamber (IC) in 
Helium  and Helium-Hydrogen at various beam intensities (noted
in units of protons/spill).    
The vertical axis is the ratio of charge collected from the IC
to the beam intensity measured by the toroid.  Each point is the average
of 10-20 beam spills.}
\label{fig_hv2mm}
\end{figure}

Tests of the chambers consisted of two complementary measurements.
The first held the beam intensity constant, while varying the voltages applied to 
the chambers.  The second held the applied voltages constant while adjusting
the beam intensity.

The results of the first test are displayed in Figures~\ref{fig_hv1mm} 
and~\ref{fig_hv2mm}.  Each chamber is exposed
to several beam intensities and the voltage varied 0-350~V.  The ratio of collected charge 
to the measured beam intensity is plotted as a function of applied
voltage. 

The ideal voltage plateau curve would consist of a quick rise to a constant 
charge collected per proton, 
independent of voltage and intensity. This constant charge collected 
would be equal to the amount of charge liberated in the gas per proton.  
At higher voltages gas amplification is expected and charge collected per proton
would increase above the plateau.  From the height of the plateau on the 1~mm chamber one 
may infer a charge ionized per proton of 2.6~pC/10$^7$ protons, or 1.6~electron-ion
pairs per proton incident on a 1~mm gas gap.  

\begin{figure}[t]
\centering
\includegraphics[width=3.0in]{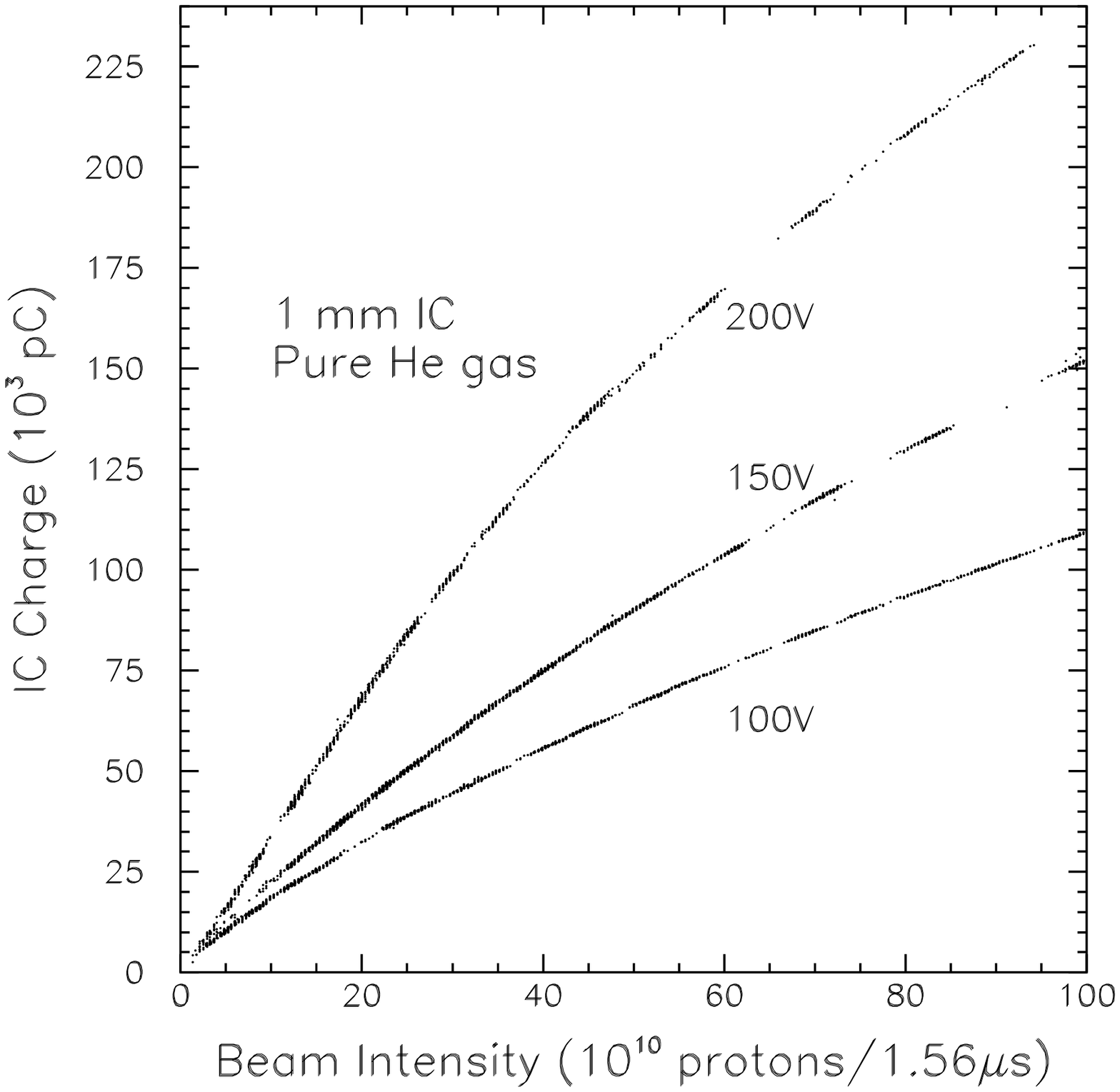}
\caption{Beam intensity scans of the 1~mm ion chamber in Helium at
various voltages.}
\label{fig_int1mmhe}
\end{figure}

The drifting ions and electrons inside the chamber establish their own 
electric fields
and the relatively slow drift velocities of the ions create a net 
space charge in the gas.  This
space charge induced electric field screens the electrodes, slowing the transit
of ions and electrons across the electrode gap.  As discussed extensively in  
\cite{Wilkinson,Rossi,Knoll}, operating ion chambers
at very large particle fluences modifies the voltage plateau curve discussed above.
If the speed of the charges is sufficiently slowed, recombination may take place.
This recombination loss is especially evident $<$150~V, 
where the lower voltages result in slower ion drift velocities and longer 
ion transit times.  This loss also increases at higher beam intensity or in the
larger 2~mm gap chamber, where space charge buildup should be worse.

Space charge effects were evident at all of the intensities delivered by the 
Booster. In Figure~\ref{fig_hv1mm} all of the curves, 
except possibly one, show a slope in the plateau region,
suggesting that there is little or no region where
the charge is collected without loss or gain.  The only useful voltage looks to be
130-190~V for the 1~mm chamber, where the curves all intersect.  
In Section~\ref{sec:sim} we analyze this in terms
of competing recombination and multiplication effects in the gas.

\begin{figure}[t]
\centering
\includegraphics[width=3.6in]{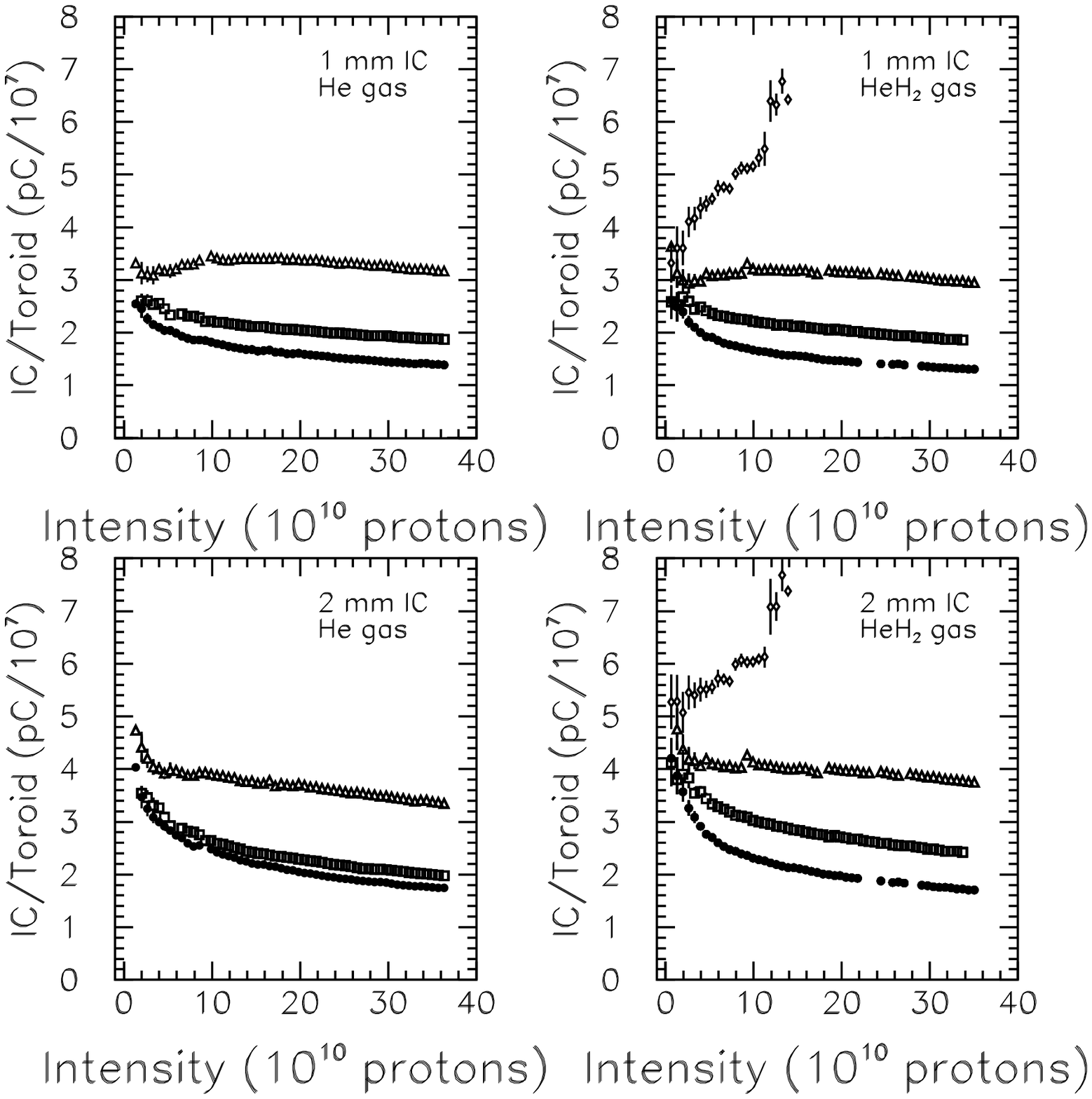}
\caption{Normalized beam intensity scans of the 1mm ion chamber at
various voltages. Here, the vertical axis is the ratio of charge collected to the
beam intensity as measured by the toroid. The points are for applied bias 
potential of 100~V $\bullet$,
150~V (125~V for 2~mm-He) $\square$, 200~V $\triangle$, and 250~V $\diamond$.}
\label{fig_nint}
\end{figure}

Figure \ref{fig_int1mmhe} displays the results of the second test performed in 
our beam test for the case of the 1~mm chamber in Helium.
The charge collected from the ion chamber is plotted as a function of beam intensity 
for several applied voltages.  Here, ideal curves would all join at low intensity.
Recombination loss, especially visible for lower voltages, cause the curves to
fall off at higher intensity.  The linear region is the operating range for the 
chamber where charge collected is proportional to incident flux.

An analysis of the 200~V Helium curve up to $20\times10^{10}$protons/spill gives a 
good linear fit ($\chi^2$/N$_{\rm DOF}$ = 456/625).  However, the intercept is less 
than zero: (-0.05 $\pm$ 0.012) $\times10^3$~pC. 
The other data could not be satisfactorily fit to a
linear dependence for beam intensities $>10^{10}$ protons/spill.

The nonzero intercept in the 200 V data is more 
apparent when the curve is plotted as a ratio, as in 
Figure~\ref{fig_nint} where the ratio of charge collected to beam intensity is 
plotted versus intensity.  Here a straight line with zero intercept in the 
previous plot corresponds to a horizontal line.  In Figure~\ref{fig_nint} 
the 200~V curve is obviously curved upward then downward as the intensity is 
increased from 1 to $20\times10^{10}$ protons/spill.  This rise is responsible for
the negative intercept in the linear fit.  In Section~\ref{sec:sim} we discuss how 
the interplay of multiplication and recombination can cause such an effect.  
Similar to the voltage plateau data, the data in Figure~\ref{fig_nint}
indicate a collected charge of $\sim$3pC/10$^7$protons in the 1~mm chamber, or 
18~electron-ion pairs per cm in the He.

The similarity between the Helium and Helium-Hydrogen data is interesting.  
Gas additives with lower excitation potentials are commonly used to increase the
drift velocity of electrons \cite{Rossi}, which might naively be expected to reduce
recombination losses.  However, the recombination loss is not substantially changed 
by the addition of H$_2$, and in Section~\ref{sec:sim} we discuss how the ion drift 
more strongly affects space charge build-up, hence recombination loss, in the chamber.

Another parameter that might be expected to vary as a result of the additive is
the ionization per proton.  The incident protons excite metastable
states in the Helium, which are followed by collision of Helium atoms with the H$_2$ additive,
yielding strong ionization of the H$_2$ \cite{Loeb}.
That the observed ionization/proton is not significantly increased by the H$_2$
additive is perhaps indicative that even the minute impurities ($<$1.5~p.p.m.)
in the nominally pure Helium gas are sufficient to increase the ionization.  
The observed 1.6-1.8~ionizations/mm per proton in our data
is notably higher than the 0.8-1.0~ionizations/mm 
inferref from the dE/dx of fast charged particles and the $w$=42~eV/ionization
\cite{ICRU} for the 
purest Helium.  Thus, the nominally pure Helium used in typical chamber applications 
is likely effectively doped.

\section{Simulation}
\label{sec:sim}

In this section we model the data collected at the Booster beam test
using a computer simulation of the pulse development in an ion chamber.  The
simulation incorporates the known drift of electrons and of Helium ions in
electric fields, the effects of volume recombination of charges, and of gas 
amplification.  The simulation follows the charges during and after the beam
spill, recording the net charge collected at the chamber electrodes.  The
simulations indicate a complex interplay between gas amplification and 
charge recombination as a result of space charge build up.  This
calculation will then be used to extrapolate to the particle fluxes and 8.6$\mu$s 
spill duration expected in NuMI.  Only Helium gas was simulated as literature 
data for the gas properties of Helium-Hydrogen are minimal, and the beamtest
data were not significantly different for the two gases

\subsection{Simulation Method}

The differential equations governing the charge flux and electric field evolution
inside the ion chamber are nonlinear.  Some earlier work analyzed special cases.
\cite{Boag} demonstrated that space charge accumulation in a continuously-ionized
chamber causes the formation of
a ``dead region'' of no electric field, and makes the approximation that all charges 
in the dead region are lost due to recombination.  In \cite{NA48}, an ion chamber
ionized over a short duration is studied, but again the assumption is made that all 
charges in the dead zone are lost to recombination.
That assumption is not valid for our experiment where the beam
pulse is of short duration allowing the dead zone to disappear before the charge 
completely recombines.

\begin{figure}[t]
\centering
\includegraphics[width=3.0in]{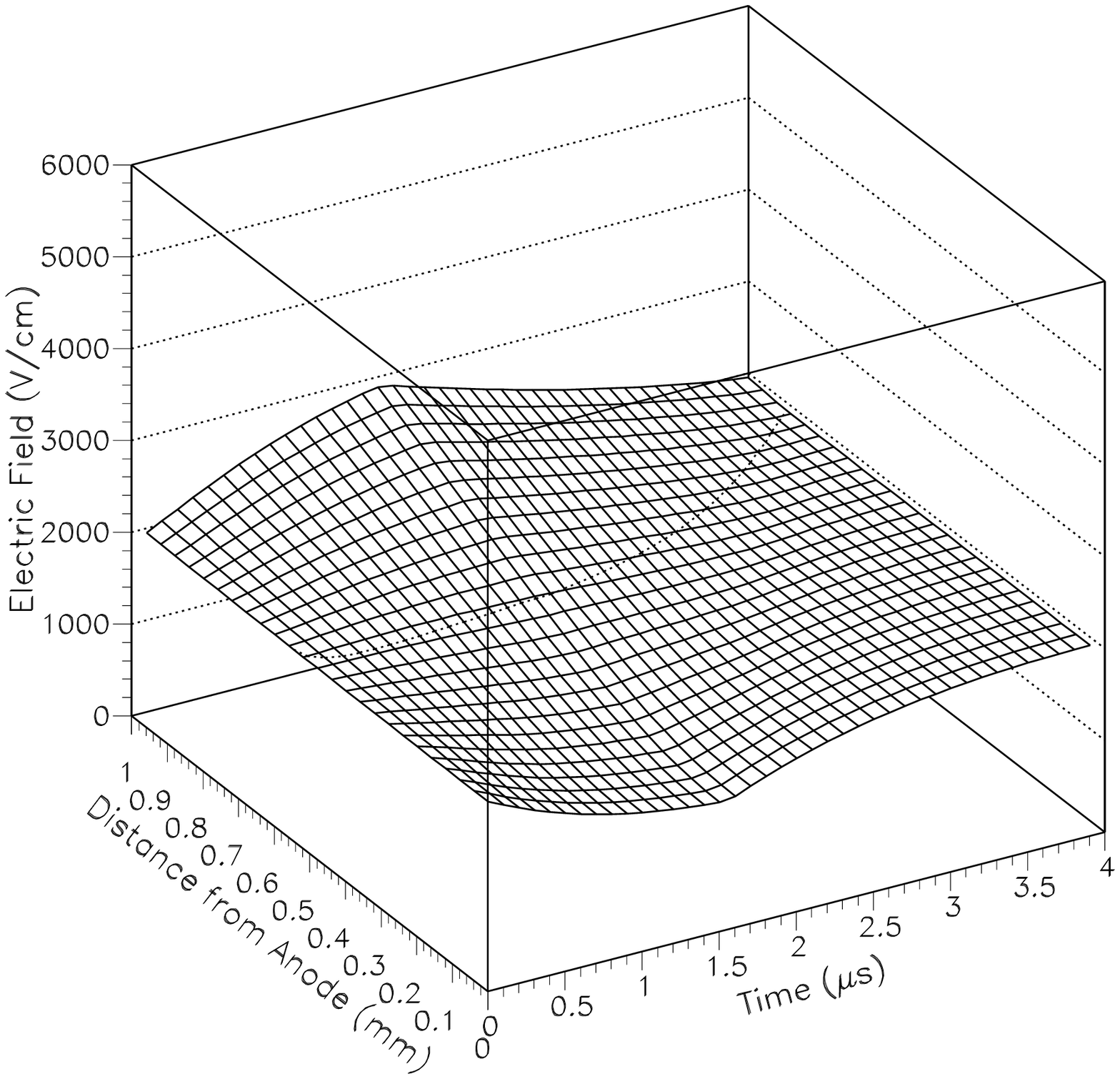}
\caption{Simulated electric field evolution in terms
of position and time for a 1~mm ion chamber operated at 200~V with ionization
of 10$^{10}$~ion./cc/$\mu$s for 1.6~$\mu$s.
Small electric field disruptions due to space charge accumulation
occur as the beam pulse develops.}
\label{fig_1mm1e10}
\end{figure}

The present work considers the time dependant case where ionization is delivered in 
short pulse of 1.6~$\mu$s and allowed to drift out of the chamber.   Our simulation is
a finite element calculation of one spatial dimension, such that there are series of 
infinite planes of charge between the electrodes.  The electrons and ions
are drifted with the velocites discussed in Section~\ref{sec:sim_gas}.  Space charge
is calculated at each step and an image charge is induced such that the potential
difference betweeen the electrodes is maintained at the applied voltage.  

\begin{figure}[t]
\centering
\includegraphics[width=3.0in]{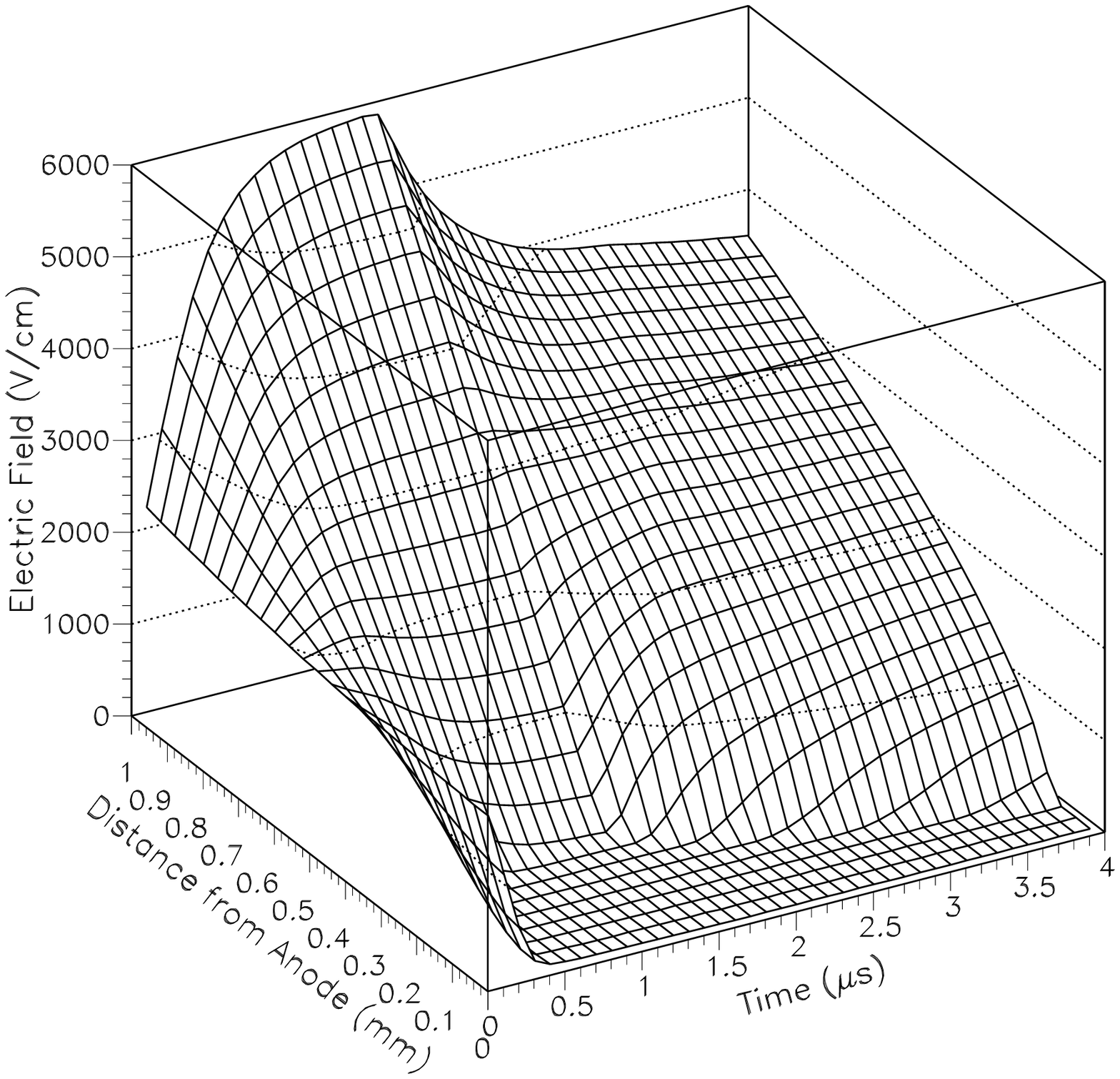}
\caption{Simulated electric field evolution in terms
of position and time for a 1~mm ion chamber operated at 200~V with ionization 
of 10$^{11}$~ion./cc/$\mu$s for 1.6~$\mu$s.  A ``dead zone'' region with
a very low electric field forms at about 0.3~$\mu$s into the beam spill, and
eventually grows to cover a third of the chamber.  The space charge accumulation
also increase the electric field in part of the chamber to the point where the 
maximum electric field at the cathode is almost three times the applied electric
field.}
\label{fig_1mm1e11}
\end{figure}

\begin{figure}[t]
\centering
\includegraphics[width=3.0in]{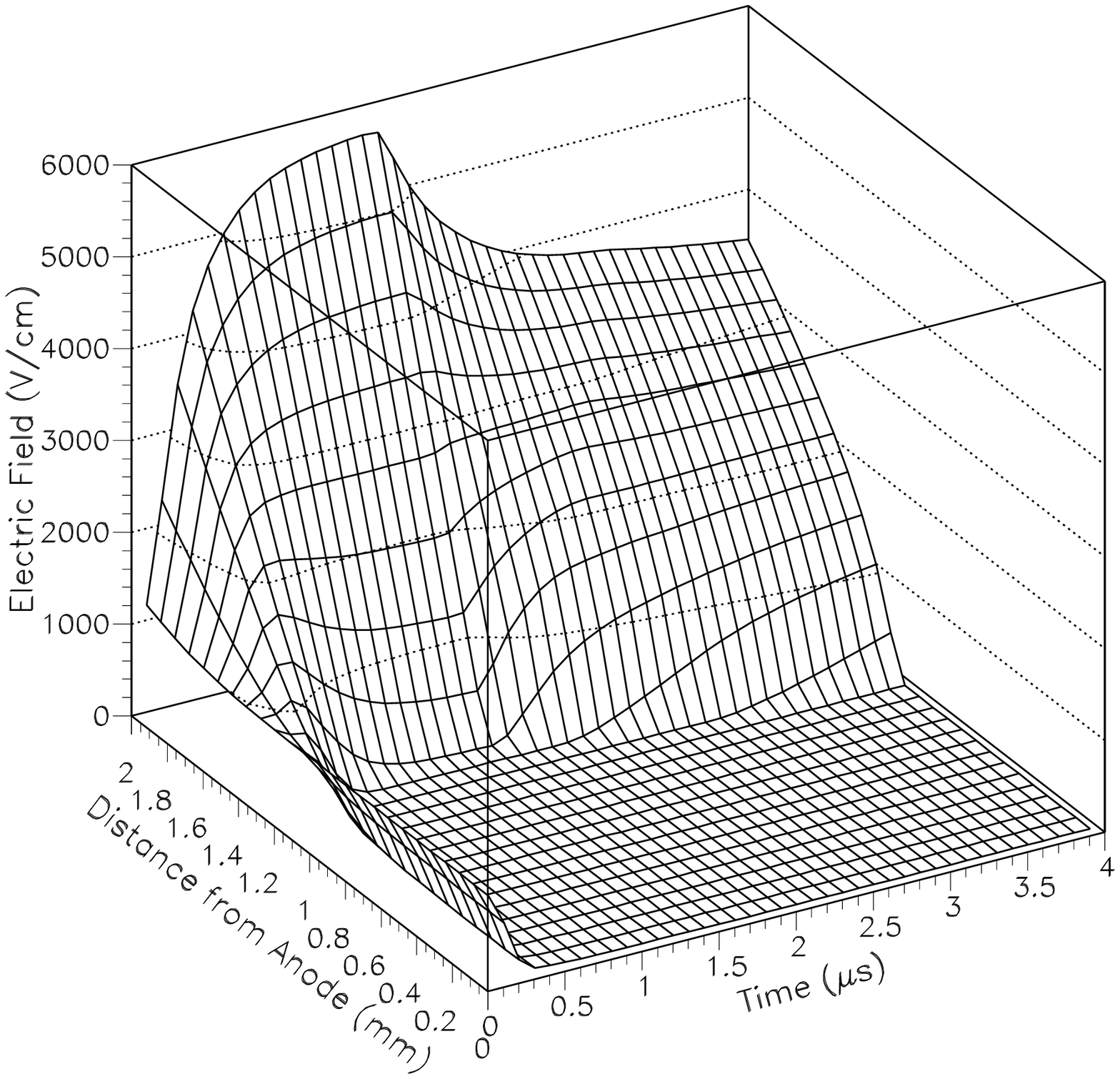}
\caption{Simulated electric field evolution in terms
of position and time for a 2~mm ion chamber operated at 200~V with ionization 
of 10$^{11}$~ion./cc/$\mu$s for 1.6~$\mu$s.  A ``dead zone'' region with
a very low electric field forms at about 0.3~$\mu$s into the beam spill, and
eventually grows to cover most of the chamber.  Shortly into the spill the behavior
of the half
of the chamber adjacent to the cathode is almost identical to that of the 1~mm chamber 
in Figure~\ref{fig_1mm1e11}.}
\label{fig_2mm1e11}
\end{figure}

Simulated electric field distributions as a function of time are shown in 
Figures~\ref{fig_1mm1e10}, \ref{fig_1mm1e11}, and 
\ref{fig_2mm1e11}.  The figures show the field development using only the 
charge transport, and ignore charge recombination or multiplication.  

The
electric field simulated in Figure~\ref{fig_1mm1e10} is calculated at 
an intensity of 10$^{10}$ ionizations/cm$^3$/$\mu$s, 
where the excess of ions slightly 
warps the field.  In Figure~\ref{fig_1mm1e11}, the ion excess is much greater, to 
the point where it entirely screens the anode from the applied field, creating a dead zone.  
As ionization continues in the dead zone the ion and electron densities there
increase because the charges are effectively trapped.  When the beam 
spill ends the dead zone slowly fades away.  Charge is able to escape from the edge
of the dead zone because the dead zone is not created by the charge inside, but 
the excess of ions outside of it.  

Figure~\ref{fig_2mm1e11} is calculated at the same
intensity as Figure~\ref{fig_1mm1e11}, but for the 2~mm gap chamber. The half of the chamber
close to the anode has almost exactly the behavior of the 1~mm case.  This similarity
is a result of the complete screening of the electric fied by space charge which makes
the rest of the chamber in the dead zone irrelevant, until the charge has a chance
to excape after the ionization period (the beam spill) ends.

\subsection{Gas Properties}
\label{sec:sim_gas}

The parameters of charge drift, loss, and amplification used in this simulation
have been taken from several literature sources.
The electron drift velocities as a function of electric field are taken from 
\cite{Dutton}.  For the drift of Helium, we use the mobility for
the He$_2^+$ ion, $\mu_{He_2^+}=20$~cm$^2$/(V$\cdot$s) \cite{Dutton}, following the 
discussion of \cite{Loeb} which notes that above pressures of a few Torr, He$^+$
ions tend to collide in the gas and form molecular ions.  Previous work 
\cite{Sauli,Blum} have cited 10~cm$^2$/(V$\cdot$s) as the relevant mobility for He.

We simulated charge loss through two- and three-body volume 
recombination of electrons ions.
\cite{Dutton}  and \cite{Brown} parameterize recombination of the form:
$dn_{-}/dt=dn_{+}/dt=- rn_{-}n_{+}$, with $n_+$ ($n_-$) the positive (negative) 
ion densities in the gas, and the measured recombination coefficient 
$r=2.4\times10^{-8}$~cm$^3$/(ion$\cdot$s).

We did not simulate charge loss through the process of electron attachment
to electronegative impurities and subsequent recombination.
Using the data from \cite{Wilkinson} and \cite{Brown} we find that 
the attachment of electrons to Oxygen in the gas is
insignificant for impurity levels less than 30~p.p.m. (to be compared with
1.5~p.p.m. observed in our chambers).  Attachment is
negligle over the drift region and occurs primarily in the dead zone
where the electrons persist for a greater time.  There, the electrons
will attach with a time constant $\tau=(67/x)$ ms, where $x$ is the
impurity level in p.p.m.  The electrons persist in the dead zone
time of order 10 $\mu$s and the impurities in our beam test were of order 1 p.p.m
giving about 0.1\% attachment.

Charge amplification in the chamber gas is modeled via $dN/dx=N\alpha$, with the 
Townsend coefficient $\alpha$ modelled via $\alpha/P=A\exp[-B/(X/P)]$, where
$P$ is the chamber gas pressure, $X$ is the applied electric field, and the 
parameters $A=3$~ion~pairs/(cm-Torr) and $B=25$~V/(cm-Torr) \cite{vonEngel}.  
This parameterization is taken from a source with similar gas purity, which is 
of importance because the purest Helium gas has significantly lower Townsend 
coefficient but is sensitive to even a few p.p.b. impurity level \cite{Dutton}.

The ionization rate in the simulation is quoted in units of 
ionizations/cm$^3$/$\mu$s in order to factor out the question of how many 
ionizations are created by an incident proton.  Given $dE/dx$ and $W$, the 
average energy loss necessary to create one ionization in the gas,
for Helium, values of 8-16/cm have been derived \cite{ICRU,Sauli,Blum,PDG},
but this too is sensitive to impurity level.  For comparison to our 
data from the Booster beamtest, 1~ion./cc/$\mu$s corresponds to 
0.6~protons/spill if the beam spot size of 5~cm$^2$ and 
$\sim$16~ionizations/cm in Helium are used.

\begin{figure}[t]
\centering
\includegraphics[width=3.2in]{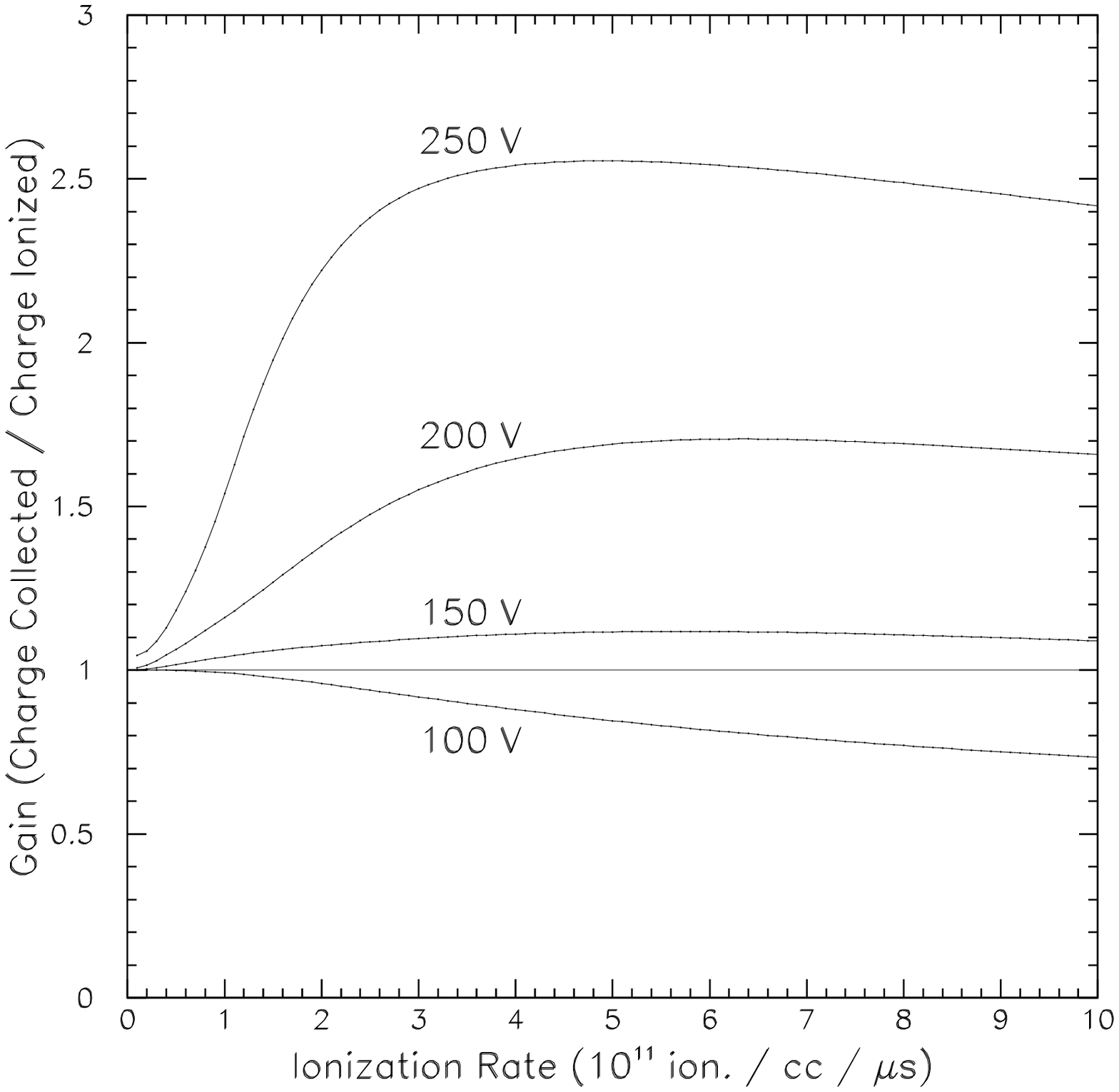}
\caption{Simulated normalized intensity scans of a 1~mm 
in chamber operated at various voltages where charge recombination and
gas multiplication are included.  The vertical axis is the ratio
of charge collected to the amount of charge initially ionized in the chamber
The horizontal line at 1.0 on the vertical axis is expected when the 
space charge effects of recombination and multiplication are ignored.
\label{fig_multnint}}
\end{figure}

\begin{figure}[t]
\centering
\includegraphics[width=3.2in]{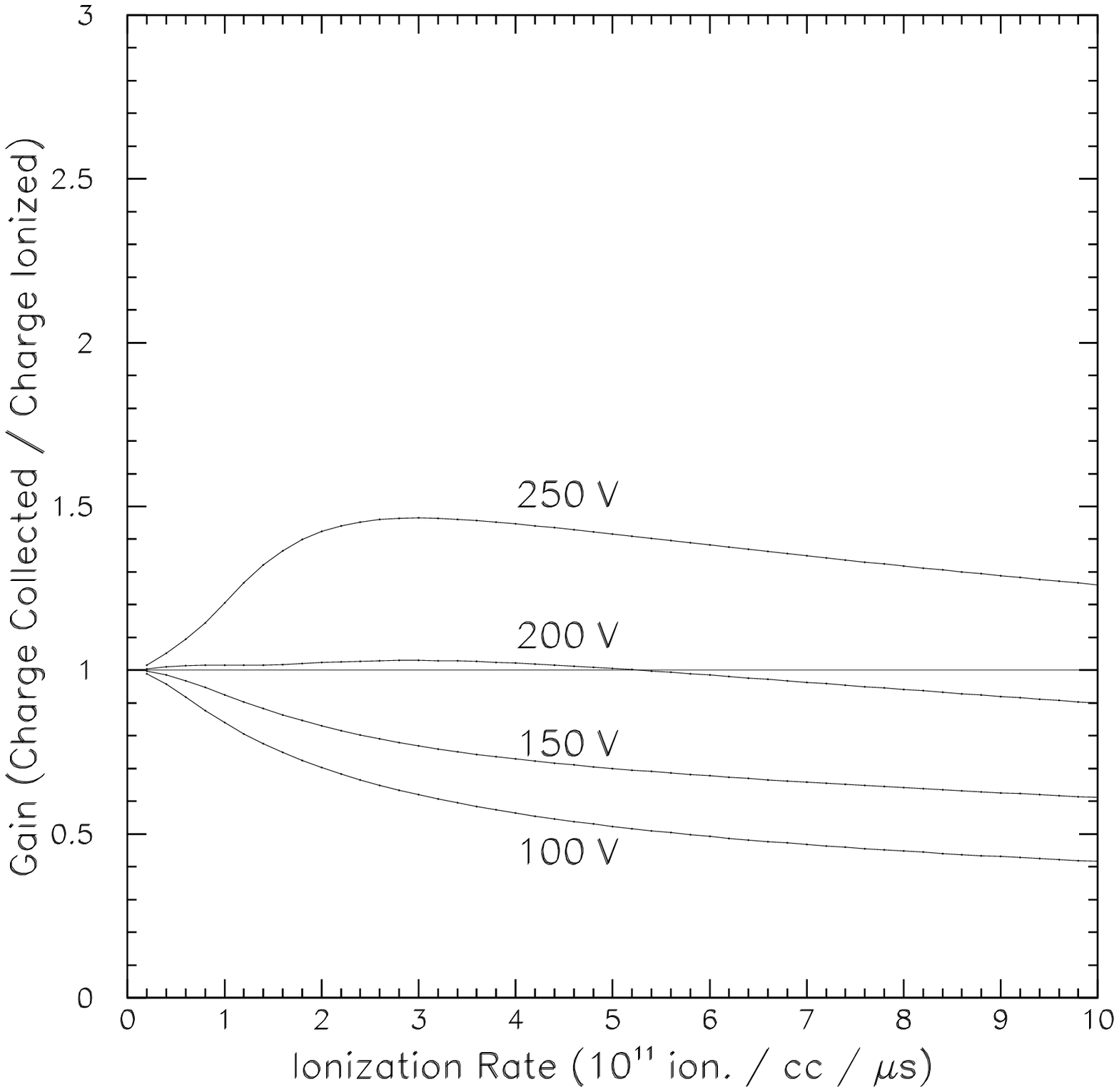}
\caption{Simulated normalized intensity scans of a 2~mm 
in chamber operated at various voltages where charge recombination and
gas multiplication are included.  The vertical axis is the ratio
of charge collected to the amount of charge initially ionized in the chamber
The horizontal line at 1.0 on the vertical axis is expected when the 
space charge effects of recombination and multiplication are ignored.
\label{fig_multnint2}}
\end{figure}

\subsection{Simulation Results}

\begin{figure}[t]
\centering
\includegraphics[width=3.2in]{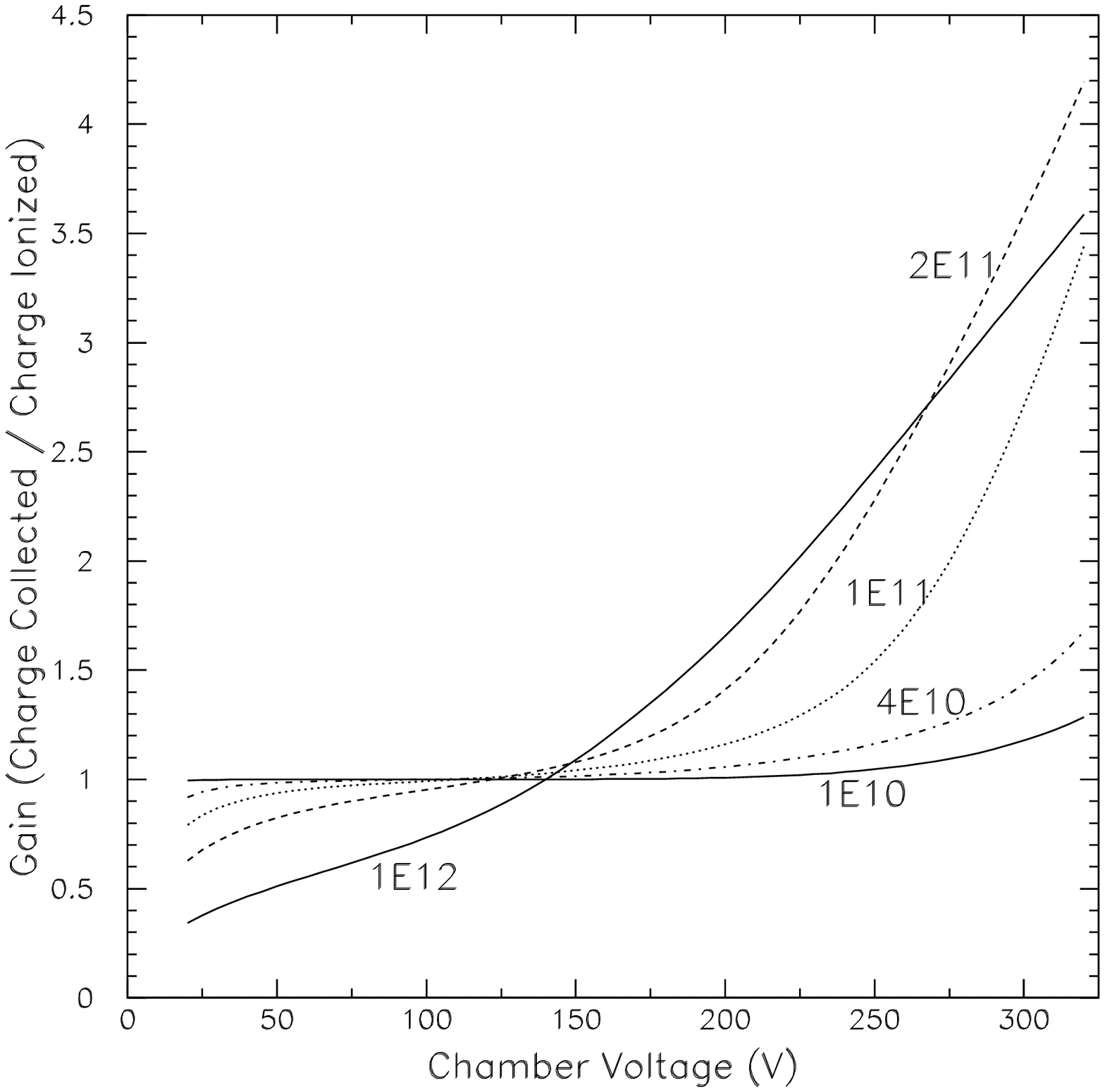}
\caption{Simulated voltage scans of a 1~mm 
ion chamber operated at various intensities where charge recombination 
and gas multiplication are included. The vertial axis is the ratio
of charge collected to the amount of charge initially ionized in the chamber.
\label{fig_multvolt}}
\end{figure}

\begin{figure}[t]
\centering
\includegraphics[width=3.2in]{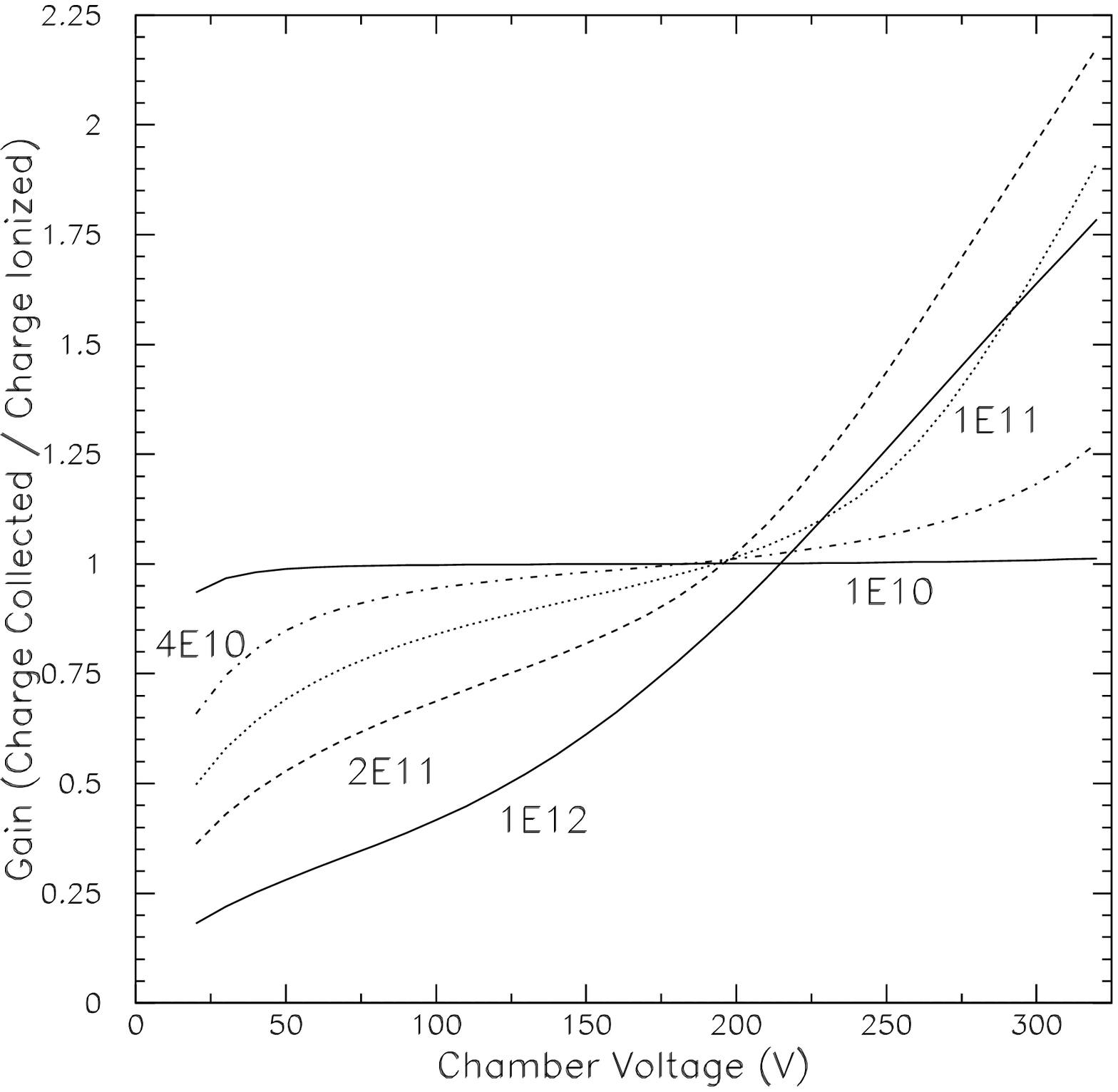}
\caption{Simulated voltage scans of a 2~mm 
ion chamber operated at various intensities where charge recombination 
and gas multiplication are included. The vertial axis is the ratio
of charge collected to the amount of charge initially ionized in the chamber.
\label{fig_multvolt2}}
\end{figure}

We found that the simulated space charge accumulation causes not only recombination losses,
but the high fields also cause significant multiplication.  This result can be 
seen in Figures~\ref{fig_multnint} and \ref{fig_multnint2} where 
the charge collected per ionization can actually
increase with ionization, until recombination takes over.  This should be compared to 
Figure~\ref{fig_nint}.  Both the data and the simulation indicate such an effect.
This feature of the curve makes the fit line have a negative intercept,
as stated earlier.  The lower voltage required in the simulation to generate this
effect suggests that multiplication may be overstated in simulation.

The simulation was used to generate voltage plateau curves as well.  The results, 
shown in Figures~\ref{fig_multvolt} and \ref{fig_multvolt2}, can be compared to 
Figures~\ref{fig_hv1mm} and \ref{fig_hv2mm}, respectively.  The behavior
is marked by greater losses at high intensity and low 
voltage, and also by a crossing point where the curves all approach each other.  
The crossing point occurs when the charge collected equals the charge liberated in 
the gas.  In the simulation this occurs at $\approx$140~V and 210~V for the 1 and 
2~mm chambers.  In the data, these points occur at 175~V and and 190~V. 
Another feature is that higher intensities can show gain at lower voltages, resulting
in the curves diverging after the crossing point.

The behavior observed in the data is nominally reproduced by the simulation.  
The crossing point occurs at a higher voltage than in 
the simulation of the 1~mm chamber, again suggesting that multiplication 
is overstated there.
The crossing point can be moved left and right by simply changing the value 
of the Townsend coefficient $\alpha$, and is particularly sensitive to the 
parameter $B$.  Hence the data can constrain the value of $\alpha$, and 
possibly the recombination coefficient $r$.

\section{Conclusion}

We have performed a beam test of Helium- and Helium-Hydrogen-filled ionization 
chambers at the Fermilab
Booster accelerator.  We have compared the experimental results to our own 
calculation of the expected charge collection in such chambers.  While we have yet 
to extrapolate our calculations to the anticipated NuMI beam environment using 
the constraints of our beam test data, several effects of interest are observed.
First, calculation of the expected ionization per charged particle in Helium based 
$dE/dx$ loss and the value $w=42$~eV from \cite{ICRU} may ignore $\sim$p.p.m. 
impurities in most chamber gas which result in additional ionization.
Second, similar to previous work, our results indicate that 
space charge effects induce recombination losses in the collected
charge at high particle fluxes; however, our results indicate additionally
that gas amplification occurs in the ion chambers at high beam intensities
due to the large space charge build-up of the electric field.  The gas 
amplification gains compete with the losses due to recombination, effectively 
extending the range of linear response of the ion chamber with respect to beam
intensity.

\section{Acknowledgements}
We gratefully acknowledge the excellent facilities and technical help provided by the Fermilab 
Booster Radiation Damage Facility group, J. Lackey, M. Ferguson, T. Sullivan as well as 
B. Webber, Fermilab Proton Source Department Head. 
\end{document}